\documentstyle[sprocl]{article}
\input epsf
\input psfig 

\def\be{\begin{equation}}
\def\ee{\end{equation}}
\def\bea{\begin{eqnarray}}
\def\eea{\end{eqnarray}}

\begin{document}
\title{Formation of Topological Defects}

\author{Tanmay Vachaspati}
\address{
Physics Department,
Case Western Reserve University,\\
Cleveland OH 44106-7079, USA.
}

\maketitle

\abstracts{
In these lectures, I review cosmological phase transitions 
and the topological aspects of spontaneous symmetry breaking. I then
discuss the formation of walls, strings and monopoles during 
phase transitions including lattice based studies of defect formation 
and recent attempts to go beyond the lattice. The close connection 
of defect formation with percolation is described. Open problems and 
possible future directions are mentioned.
}

%
%
%
%

\section{Motivation and Introduction}

An exciting development in cosmology has been the 
realization that the universe may behave very much
like a condensed matter system. After all, the cosmos
is the arena where very high energy particle physics 
is relevant and this is described by quantum field
theory which is also the very tool used in condensed matter physics.

In condensed matter systems we have seen a rich array of
phenomenon and we expect that the cosmos has seen an equally 
rich past. A routine observation in condensed matter (and 
daily life!) is that of symmetry breaking which is 
the basis of all schemes in particle physics to achieve
unification of forces. Symmetry breaking would lead to phase 
transitions in the early universe, making their study 
crucial to our understanding of the cosmos. There are many 
cosmic theories that hinge on processes happening during 
phase transitions. These include a large number of inflationary 
models, baryogenesis and structure formation. 

Many of the concepts that have gone into the explosion of
cosmology in the last two decades are linked to
each other. As an example, consider the birth of the inflationary
idea. The success of the electroweak model led particle physicists
to attempt to unify the strong force with the electroweak forces
by postulating Grand Unified Theories (GUTs). In such theories,
a number of phase transitions occur and, based on mathematical
results on the topology of group manifolds, 
it is known that GUTs always contain topological
defects known as magnetic monopoles. The occurence of phase
transitions in cosmology then tells us that monopoles must
have formed in the early universe~\cite{twbk}. In fact,
standard cosmology then predicts
an over-abundance of magnetic monopoles in our present universe
that is clearly in conflict with observations \cite{preskill}. 
This head-on
confrontation of particle physics and cosmology led Guth \cite{guth}
to come up with the idea of an inflationary universe - an idea
that is now deeply embedded in cosmology as it not only solves
the monopole problem but also several other unrelated problems
that standard (pre-inflationary) cosmology did not address.

Magnetic monopoles are one example of topological defects that
can arise during phase transitions. But topological defects
can occur in other varieties as well. They can be one dimensional,
in which case they are called ``strings''. If they are two dimensional,
they are called ``domain walls''. Magnetic monopoles and domain
walls are two kinds of topological defects that 
have unpleasant cosmological consequences unless they are kept very
light or are somehow eliminated at some early epoch (for example, by
inflation). Cosmic strings are believed to be more benevolent
and may even have been responsible for structure formation in the
universe if they were formed at the GUT phase transition. 

The study of topological defects is fascinating for several reasons.
As in the example of magnetic monopoles and inflation above, they can 
provide important constraints on particle physics models and cosmology.
On the other hand,
if a GUT topological defect is found, it would provide a direct window
on the universe at about $10^{-35}$ secs. In a manner of speaking, a
part of the early universe is trapped in the interior of the defect -
much like dinosaur fossils trapped in amber. GUT
topological defects would also shed light on the problem of how
galaxies were formed. Their discovery would give us important
constraints on the symmetry structure of very high energy particle 
physics - this is non-perturbative information. In addition, it would
give us information about phase transitions in the early universe,
giving confidence in our understanding of the thermal history of the
universe. Also, the topological defects would most likely have 
tremendous astrophysical impact since they are generally very massive 
and have unusual gravitational and electromagnetic properties. 

In the early 80's, people would talk of the marriage of particle
physics and cosmology. I actually think there is something wrong
with this picture since a third party is also involved. This is 
condensed matter physics. When it comes to understanding cosmological
phase transitions, we are forced to consider the corresponding
advances in condensed matter physics since many of the ideas are the same. 
Topological defects have been studied by condensed matter physicists 
for many, many years. If we want experimental input in our theories 
of the very early universe, we must go to the condensed matter laboratory. 
This is a growing area of research - condensed matter experiments are
being inspired by cosmological questions, and, cosmologists are 
revising their theories based on experimental input.

In these lectures, my aim is to discuss the formation of topological
defects and to bring the reader to a point where the most relevant
questions are apparent. The existing work on topological defects
paints a certain picture of their formation but there are some
limitations. I will describe both the picture and the limitations.
I will start out by describing phase transitions and the ``effective
potential'' way of 
treating them in field theory. Then I will describe why topological
defects come about and finally get to the work on their formation 
in a phase transition. It is impossible to describe every aspect
of this subject within these lectures especially since the subject
branches out into a large variety of different areas of research.
The references provided in the bibliography should help the reader
dive deeper into whichever area he/she chooses to pursue.

\section{Phase Transitions} 

\subsection{Effective Potential: Formalism}

In statistical mechanics, the basic quantity one tries to
find is the partition function $Z$ 
$$
Z = \sum e^{-\beta H}
$$
where $\beta = 1/T$ and the sum is over all states with
energy $H$. From the partition function one
can derive the Helmholtz free energy $A$ by
$$
Z = e^{-A}
$$
and the derivatives of $A$ then lead to the thermodynamic
functions. 

To discuss phase transitions, one performs a Legendre
transform on the Helmholtz free energy to get the Gibbs
free energy $G$. For a gas, we have
$$
G(P,T) = A + PV
$$
while for an ensemble of spins,
$$
G(M,T) = A + HM
$$
where $H$ is the external magnetic field and $M$ is the
magnetization of the system. 

In the magnetic system, we have
$$
{{\partial G}\over {\partial M}} = H
$$
and so, in the absence of an external magnetic field, the
minima of $G$ describe the various phases of the system.

In field theory, these concepts carry over almost word
for word \cite{peskinbook}. 
For the time being we restrict ourselves to the zero
temperature case. Then, in analogy with the partition function, 
one defines the 
generating functional in the presence of an external current $J(x)$ 
(analogous to the external magnetic field)
\begin{equation}
Z[J] = \int D\phi {\rm exp}[i\int_0^{\cal T} dt \int d^3 x 
({\cal L} [\phi ] + J(x)\phi (x) ) ] 
\end{equation}
where, the time integration is over a large but finite interval
and ${\cal L}$ is a suitable Lagrangian density for the system.
In other words,
\begin{equation}
Z[J] = <\Omega | e^{-iH{\cal T}} | \Omega > \equiv  {\rm exp}[ - E[J]] 
\end{equation}
where, $|\Omega >$ is the vacuum state and $H$ is the Hamiltonian.
The energy functional $E[J]$ is the analog of the Helmholtz
free energy. (Note however that $E$ is not really an energy - for
example, it does not have the dimensions of energy.)
Now, we find,
\begin{eqnarray*}
{{\delta E [J]} \over {\delta J(x)}} & = &
- { {\int D\phi e^{i\int(\cal L +J\phi )} \phi(x)} \over
    {\int D\phi e^{i\int(\cal L +J\phi )}        } } \\
& = &<\Omega | \phi (x) | \Omega >_J \\
& \equiv & - \phi_{cl} (x) 
\end{eqnarray*}
which is (minus) the order parameter. Next we get the effective action, 
which is the analog of the Gibbs free energy, by Legendre transforming
the energy functional:
$$
\Gamma [ \phi_{cl} ] = -E[J] - \int d^4 y J(y) \phi_{cl}(y) \ .
$$
To obtain the value of the order parameter for any given external
current, we need to solve:
$$
{{\delta \Gamma [\phi_{cl}]} \over {\delta \phi_{cl} (x)}} = - J(x) 
$$
and, in particular, when the external current vanishes, the 
phases are described by the extrema of the effective action.

So far we have taken the external current to be a function of
space but, in practice, the current is taken to be a constant
and $\phi_{cl}$ is also constant. Then, in this restricted case,
only non-derivative terms can be present in the effective action
and the effective action leads to an effective potential with
quantum corrections:
$$
V_{eff,q} (\phi_{cl}) \equiv - {1 \over {{\cal V}{\cal T}}} 
\Gamma [\phi_{cl}]
$$
where, ${\cal V}{\cal T}$ is the spacetime volume.

These issues and, in particular, the correspondence between 
statistical mechanics and field theory, are described very clearly 
in the textbooks by Peskin and Schroeder \cite{peskinbook} and
Rivers \cite{rr}.

\subsection{Effective Potential: Quantum Corrections}

The actual calculation of the quantum effective potential is done
by using perturbation theory \cite{scew,linde}. 
The general result is as follows. For the model:
$$
{\cal L} = {1\over 2} D_\mu \phi_i D^\mu \phi_i - V(\phi_i )
- {1\over 4} F^a_{\mu \nu} F^{a\mu \nu} + {\cal L}_F
$$
where
$$
\partial_\mu \phi = (\partial_\mu -ie A^a_\mu T^a )\phi
$$
$$
F^a_{\mu \nu} = \partial_\mu A^a_\nu - \partial_\nu A^a_\mu
 +e f_{abc} A^b_\mu A^c_\nu
$$
$$
{\cal L}_F = i{\bar \psi} \gamma^\mu D_\mu \psi
              - {\bar \psi} \Gamma_i \psi \phi_i
$$
in conventional notation,
the one loop correction to the potential is the Coleman-Weinberg
correction:
\begin{equation}
V_{1,q} (\phi ) = {1 \over {64 \pi^2}}
  [  {\rm Tr} (\mu^4 {\rm ln} {{\mu^2}\over {\sigma^2}}) +
    3 {\rm Tr} (M^4 {\rm ln} {{M^2}\over {\sigma^2}}) -
    4 {\rm Tr} (m^4 {\rm ln} {{m^2}\over {\sigma^2}}) ] \ .
\label{v1qgen}
\end{equation}
Here $\mu$, $M$ and $m$ are the scalar, vector and spinor masses
and $\sigma$ is a renormalization scale. The factors of 1, 3 and
4 in front of the  three terms are due to the spin degrees of freedom
of a scalar, massive vector and spinor (fermion-antifermion) field.
Also, the - sign in front of the fermionic contribution is due to 
Fermi statistics. The renormalization conditions used to derive
this form of $V_{1,q} (\phi )$ are the ``zero momentum'' conditions
\cite{rr}.

As a specific simple example, consider the Lagrangian 
\begin{equation}
L = {1\over 2} (\partial_\mu \Phi )^2  - V(\Phi )
\label{simplemodel}
\end{equation}
where, $\Phi$ is real and
$$
V(\Phi ) = - {{\mu^2} \over 2} \Phi ^2 + 
{\lambda \over 4} \Phi^4 \ .
$$
Next write 
$$
\Phi (x) = \phi_{cl} + \chi (x) 
$$
where, $\phi_{cl}^2$ is a constant and equal to
$\mu^2 /\lambda$ at tree level. Then,
$$
L= {1 \over 2} (\partial_\mu \chi )^2 - {1\over 2}
(3\lambda \phi_{cl}^2 - \mu^2 ) \chi^2 - \lambda \phi_{cl} \chi^3 -
{\lambda \over 4} \chi^4 + 
{1\over 2} \mu^2 \phi_{cl}^2 - {\lambda \over 4} \phi_{cl}^4
$$
The last two terms are the tree level
effective action for $\phi_{cl}$. We now want to find the
one loop correction to the effective action. This comes
about due to two ingredients: (i) the $\chi$ fluctuations 
contribute to the energy of the vacuum, and, (ii) the mass
of the $\chi$ particles depends on the value of $\phi_{cl}$. 
Hence the $\chi$ vacuum fluctuations (loops in Feynman
diagram language) contribute to the potential felt by 
$\phi_{cl}$.

There is only one Feynman diagram that contributes at one
loop - simply a $\chi$ particle propagating in a loop. Then
the contribution to the effective potential from this one
loop is:
$$
V_{1,q} (\phi_{cl}) = {1\over {2(2\pi )^4}} 
\int d^4k {\rm ln}[k^2 +m^2 (\phi_{cl})]
$$
where,
$$
m^2 (\phi_{cl}) = (3\lambda \phi_{cl}^2 - \mu^2 )
$$
A clear physical meaning can be obtained by performing
the integration over $k_0$:
\begin{equation}
V_{1,q} (\phi_{cl} ) = {1\over {(2\pi )^3}}
   \int d^3k [{\vec k}^2 +m^2 (\phi_{cl})]^{1/2}
\label{v1q}
\end{equation}
where an infinite constant has been removed by shifting
the zero energy level. The integrand now is the energy
of a $\chi$ particle with momentum $\vec k$ and so the
one loop correction to the effective potential is just
the energy in all the modes of $\chi$ - corresponding to 
the sum over $\hbar \omega /2$ in field theory. In elementary
applications of quantum field theory, this zero point energy
is removed by normal ordering, but here it contains a 
non-trivial dependence on $\phi_{cl}$ which cannot be
removed and should be included in the effective potential.
 
The integration in eq. (\ref{v1q}) is divergent and needs to
be renormalized. This is done by choosing a set of renormalization
conditions. Linde's choice \cite{linde} is:
$$
{{dV} \over {d\phi}} \biggr |_{\phi = \mu \lambda^{-1/2}} = 0
$$
and
$$
{{d^2V} \over {d\phi ^2}} \biggr |_{\phi = \mu \lambda^{-1/2}} = 
 2\mu^2
$$
where we have dropped subscripts on the effective potential and
the order parameter to simplify notation.
The final result for the effective potential is:
$$
V(\phi ) = -{{\mu^2}\over 2} \phi^2 + {\lambda\over 4} \phi^4
+{1\over{64\pi^2}} (3\lambda \phi^2 -\mu^2)^2 {\rm ln} 
\biggr ( {{3\lambda \phi^2 -\mu^2} \over {2\mu^2}} \biggl ) +
{{21\lambda\mu^2} \over {64\pi^2}}\phi^2 -
{{27\lambda^2}\over{128\pi^2}} \phi^4 \ .
$$
This result differs from the general form in eq. (\ref{v1qgen})
because of a different choice of renormalization conditions.
The physical consequences are, however, independent of the
renormalization conditions one chooses \cite{linde}.

\subsection{Effective Potential: Thermal Corrections}

Now we look at thermal corrections to the potential. 
These corrections can
be understood as follows: a thermal bath necessarily contains
a thermal distribution of particles. The properties of any
particle is influenced by its interactions with the particles
in the thermal background. This leads to an effective Lagrangian 
in which the effects of temperature are already included.

The derivation of the temperature dependence of the effective 
potential is closely analogous to the derivation of the quantum
corrections. The calculation is now done in Euclidean space with
the Euclidean time coordinate being periodic with period 
$1/(2\pi T)$.  This means that the zeroth component, $k_0$, of the
four momentum of a particle is now discrete. For bosons, periodic
boundary conditions are used and so $k_0$ is replaced by $n(2\pi T)$
while for fermions, anti-periodic boundary conditions are necessary
and $k_0$ is replaced by $(n+1/2)2\pi T$ where $n$ is
any integer. Then integrals over $k_0$ get transformed to a sum over 
$n$:
So
$$
\int dk_0 \rightarrow 2\pi T \sum_{n=-\infty}^{n=+\infty}
$$

For example, in the simple model at the end of the previous
section, the temperature dependent correction to the effective
potential is:
$$
V_{1,T} (\phi) = {T\over {2(2\pi )^3}}
\sum_{n=-\infty}^{n=+\infty} \int d^3 k 
{\rm ln}[(2\pi n T)^2 + {\vec k}^2 +m^2 (\phi )]
$$
where, once again,
$$
m^2 (\phi ) = 3\lambda \phi^2 - \mu^2 \ .
$$
The result of doing the sum and the integration and applying
the renormalization conditions is:
\begin{equation}
V(\phi ,T) = -{{\mu^2}\over 2} \phi^2 + {\lambda\over 4} \phi^4
-{{\pi^2}\over {90}} T^4 + 
{{m^2(\phi )} \over {24}} T^2
\label{vpt}
\end{equation}
in the temperature range $T >> m$ to first order in $\lambda$.

We now describe a more intuitive (but equivalent) way of 
deriving the thermal corrections to the potential. The idea 
is that if we write 
$$
\Phi = \phi +\chi
$$
there is a thermal background of $\chi$ particles that
contribute to the effective potential for $\phi$.
This contribution can be calculated by taking the ensemble
average of the microscopic (bare) potential for $\phi$.
So,
$$
V_{eff, T} (\phi ) = -{\mu^2 \over 2} < (\phi +\chi )^2>
           +{\lambda \over 4} < (\phi +\chi )^4 >
$$
where $<>$ denotes ensemble averaging. The ensemble average
of odd powers of $\chi$ vanish by symmetry while the average
of $\mu^2 \chi^2$ and $\chi^4$ simply shift the zero level. 
The only non-trivial contribution comes from
$$
{{3\lambda} \over 2} \phi^2 <\chi^2 > \ .
$$
Replacing $\chi$ by an expansion in terms of creation and
annihilation operators gives:
$$
<\chi^2 > = {1 \over {(2\pi )^3}} \int 
             {{d^3 k} \over {2\omega_k}}<2 a_k^{\dag} a_k +1>
$$
The operator $a_k^{\dag} a_k$ is simply the number operator
and, since the number distribution of spin zero particles 
with momentum $\vec k$ in a thermal bath is given by the Bose 
distribution, we have
$$
< a_k^{\dag} a_k > = n(\omega_k ) = ( {e^{\omega_k /T} -1} )^{-1}
$$
where, $\omega_k = ({\vec k}^2 + m^2)^{1/2}$.
After subtracting out the constant infinite piece from $<\chi^2 >$
we find
$$
V_{eff, T} (\phi ) =  \biggl [ -{{\mu^2}\over 2} + 
{{3\lambda T^2} \over {4\pi^2} } \int 
{{k^2 dk} \over {\sqrt{k^2+{\bar m}^2}}} 
{1 \over { e^{{\sqrt{k^2+{\bar m}^2}}/T} -1 }}
\biggr ] \phi^2  +{\lambda\over 4}\phi^4
$$
where we have also rescaled the integration variable by $1/T$ to
make it dimensionless and ${\bar m}=m/T$. In the limit of large
$T$, we have ${\bar m} \rightarrow 0$ and the integral can be
done in closed form. In this limit, the result
reduces to eq. (\ref{vpt}) without the $\phi$ independent terms.

\begin{figure}[tbp]
\centerline{
\epsfxsize = 4 in 
\epsfbox{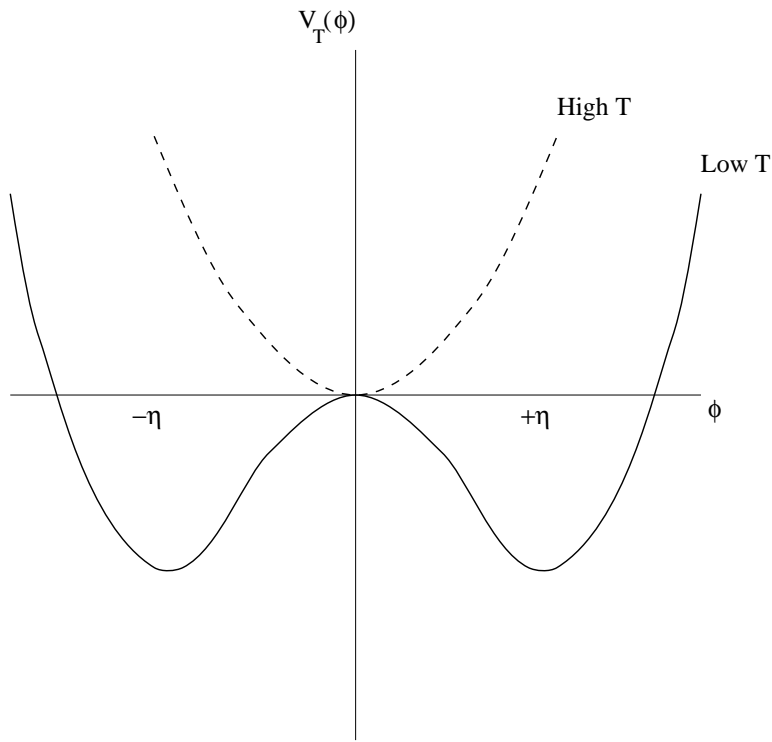}
}
\caption{Effective Potential.\label{fig:effpot}}
\end{figure}

A plot of the effective potential is shown in Fig. \ref{fig:effpot}.
At low temperatures, there are two minima 
and, in a rapid quench, the system would have
to transit from the minimum at $\phi = 0$ to the
minimum at $\phi \ne 0$.  Here the transition can take
place continuously and is a second-order phase transition.

A number of different behaviours for the effective potential
have been found. In certain systems, there are two
phases with an energy barrier separating them. This leads
to first order phase transitions. If the barrier separating
the two phases is very large, the
transition from one phase to another would have to be somehow
activated over the barrier or quantum mechanical tunneling
would eventually complete the transition. However, both
processes can be very slow to occur and so the system
can be trapped in the higher energy phase (false vacuum) for a 
long time. In other words, the system can supercool.

Note that we have been working with examples where cooling
leads to spontaneous reduction in symmetry. This is indeed
generic. However, many instances are also known where cooling
leads to symmetry restoration \cite{sw1}. This has been
observed in Rochelle salts.

\subsection{Order of the Transition: Ehrenfest Classification}

The role of the thermodynamic potentials is played by 
the effective potential. So one should take derivatives
of $V_{eff}$ with respect to the temperature and find the 
order of the lowest derivative which is discontinuous
at the critical temperature.
This will give the order of the phase transition.

As an example, consider the simple model in eq. (\ref{simplemodel}) 
for which the effective potential is given in eq. (\ref{vpt}). For
convenience, let us write:
$$
V (\phi ; T ) = -{1\over 2} M^2 (T) \phi^2 + {\lambda\over 4}\phi^4
$$
where, we have defined $M^2$
to absorb the lengthy expressions in eq. (\ref{vpt}).
The critical temperature, $T_c$, is defined by
$$
M^2 (T_c )=0 \ .
$$
The stable phases of the system are defined by 
$$
{{dV}\over {d\phi }} = 0
$$
and with the second derivative being positive.
Hence, when $M^2(T)$ is negative
({\it i.e.} high temperatures), the phase is:
$$
\phi = 0 \ , \ \ \ {\rm Phase \ I}
$$
and when $M^2(T)$ is positive ({\it i.e.} low temperatures):
$$
\phi = {{M(T)} \over {\sqrt{\lambda}}} \ , \ \ \ {\rm Phase \ II}
$$
Now we find the potential in both phases:
$$
V_I (T) = 0 \ ,
$$
$$
V_{II} (T) = - {{[M(T)]^4}\over {4\lambda}} \ .
$$
From these expressions, using $M(T_c)=0$, we have
$$
V_I (T_c) = 0 = V_{II}(T_c) \ .
$$
Also,
$$
{{dV_I} \over {dT}} \biggr |_{T_c} = 0 = 
{{dV_{II}} \over {dT}} \biggr |_{T_c}
$$
since $M^2 (T)$ is of the form $(\mu^2 - aT^2)$ where
$a$ is a constant. (Therefore the derivative of $M^2 (T)$
with respect to $T$ is well behaved and so the derivative
of $M^4$ vanishes.) However,
$$
{{d^2V_I} \over {dT^2}} \biggr |_{T_c} = 0 \ne 
{{d^2V_{II}} \over {dT^2}} \biggr |_{T_c}
$$
and so we conclude that the phase transition is second order.

The above scheme (Ehrenfest classification) for defining the order of 
a transition in terms of discontinuities in the derivatives of the potentials 
can fail if any of the derivatives of the 
potential do not exist. Generally, one says that the phase transition 
is first order if there is a barrier in the effective potential
that separates the two vacuua. If the tunneling probability
from the false vacuum to the true vacuum is very small, the 
phase transition is said to be strongly first order but if the 
tunneling rate is not too small, it is weakly first order. If there
is no barrier between the different phases, the phase transition
is said to be second-order. However, this terminology does not
imply a definite understanding of how the phase transition proceeds.
Only in the case of a strongly first order phase transition 
does one know that the phase transition proceeds by the growth
of bubbles of the lower energy (true vacuum) phase.

\subsection{Limitations of the Effective Potential}

Even in thermodynamics, the Van der Waal's equation of
state for a gas leads to a $PV$ diagram in which the 
derivative of $P$ with respect to $V$ is positive. But
this is unphysical since it is an unstable situation:
an increase in the volume leads to an increase in the
pressure, which leads to a further increase in the
volume and so on. The resolution was found in the
assumption used to derive the $PV$ diagram - 
that the sample is entirely in
one phase. More accurately, there will be regions
of $PV$ space where the system will consist of an
admixture of phases. And the free energy of the
coexisting phases can be lower than the free energy
of just one phase. 

In our discussion of phase transitions in field theory 
we have also treated the order parameter $\phi_{cl}$ as 
being uniform in space. In reality, $\phi_{cl}$ will
vary over space and the
two phases will coexist. If one applies a Maxwell
construction to the effective potential, the result
is a straight line joining the two vacuua. 

The dynamics of transiting from one phase to the other
depends on the rate at which external parameters are
varied, the shape and structure of the effective potential,
and other factors. Gravitational effects may be important in 
certain situations too. For example, if the tunneling 
rate is very small, the universe could start inflating and the 
phase transition would never complete. To add to these complications 
is the presence of topological defects that prevent the transition
from false to true vacuua from occurring globally.

\section{Topological Defects}

We first consider the existence of topological defects.
In the context of the previous lectures, these may be
viewed as obstructions to the completion of a phase
transition. In this lecture, however, we will simply
view them as classical solutions in a model that exist 
for topological reasons. I will start out by providing
the simplest examples of topological defects and then
later discuss their classification via homotopy groups.

\subsection{Domain Walls}

Consider the $Z_2$ Lagrangian in 1+1 dimensions
\begin{equation}
L = (\partial _\mu \phi )^2 - {\lambda\over 4} (\phi^2 -\eta^2 )^2
\label{z2model}
\end{equation}
where $\phi$ is a real scalar field - also called the
order parameter. The Lagrangian
is invariant under $\phi \rightarrow -\phi$ and
hence possesses a $Z_2$ symmetry. For this
reason, the potential has two minima: $\phi = \pm \eta$.
And the ``vacuum manifold'' has two-fold degeneracy.

Consider the possibility that $\phi = +\eta$ at
$x= +\infty$ and $\phi = -\eta$ at $x=-\infty$. In
this case, the continuous function $\phi (x)$ has
to go from $-\eta$ to $+\eta$ as $x$ is taken from
$-\infty$ to $+\infty$ and so must necessarily
pass through $\phi =0$. But then there is energy in
this field configuration since the 
potential is non-zero when $\phi =0$. Also, this configuration 
cannot relax to either of the two vacuum configurations, say 
$\phi (x) = + \eta$, since that involves changing the 
field over an infinite volume from $-\eta$ to $+\eta$, 
which would cost an infinite amount of energy. 

Another way to see this is to notice the presence of a
conserved current:
$$
j^\mu = \epsilon^{\mu \nu} \partial_\nu \phi
$$
where $\mu, \nu =0,1$ and $\epsilon^{\mu \nu}$ is the
antisymmetric symbol in 2 dimensions. Clearly $j^\mu$
is conserved and so we have a conserved charge in the
model:
$$
Q = \int dx j^0 = \phi (+\infty ) - \phi (-\infty ) \ .
$$
For the vacuum $Q=0$ and for the configuration described
above $Q=1$. So the configuration cannot relax into the
vacuum - it is in a different topological sector.

To get the field configuration with the boundary
conditions $\phi (\pm \infty ) =\pm \eta$, one would have
to solve the field equation resulting from the Lagrangian
(\ref{z2model}). This would be a second order differential equation.
Instead, one can use the clever method first derived by
Bogomolnyi \cite{bogo} and obtain a first order differential
equation. The method uses the energy functional:
\begin{eqnarray*}
E & = & \int dx 
     [ (\partial_t \phi)^2 + (\partial_x \phi)^2 + V(\phi ) ] \\
& = &  \int dx  [ (\partial_t \phi)^2 + 
(\partial_x \phi + \sqrt{V(\phi )} ~ )^2 - 
2 \sqrt{V(\phi )}\partial_x \phi ]  \\
& = & \int dx [ (\partial_t \phi)^2 +
(\partial_x \phi + \sqrt{V(\phi )} ~ )^2 ] - 
2 \int^{\phi(+\infty )}_{\phi(-\infty )} d\phi ' \sqrt{V(\phi ' )} \\
\end{eqnarray*}
Then, for fixed values of $\phi$ at $\pm \infty$, the energy is
minimized if
$$
\partial_t \phi =0
$$
and
$$
\partial_x \phi + \sqrt{V(\phi )} = 0 \ .
$$
Furthermore, the minimum value of the energy is:
$$
E_{min} = 2 \int^{\phi(+\infty )}_{\phi(-\infty )} 
d\phi ' \sqrt{V(\phi ' )} \ .
$$
In our case,
$$
V(\phi ) = {\lambda\over 4} (\phi^2 -\eta^2 )^2
$$
which can be inserted in the above equations to get the ``kink''
solution:
$$
\phi = \eta {\rm tanh} \biggl ({{\sqrt{\lambda} \eta x}\over 2} \biggr )
$$
for which the energy is:
$$
E_{kink} = {4 \over 3} \sqrt{\lambda} \eta^3
$$
Note that the energy density is localized in the region where
$\phi$ is not in the vacuum, {\it i.e.} in a region of thickness
$2( \sqrt{\lambda} \eta )^{-1}$ around $x=0$.

We can extend the model in eq. (\ref{z2model}) to 3+1 dimensions and 
consider the case when $\phi$ only depends on $x$ but not on $y$ and 
$z$. We can still obtain the kink solution for every value of $y$ and
$z$ and so the kink solution will describe a ``domain wall'' in the 
$yz-$plane.

Notice that the existence of the domain wall only depends on the
fact that there were discrete vacuua in the theory.

\subsection{Cosmic Strings}

Consider the Lagrangian:
\begin{equation}
L = |\partial _\mu \phi |^2 - {\lambda\over 4}(|\phi |^2 -\eta^2 )^2
\label{glstring}
\end{equation}
where $\phi$ is now taken to be a complex scalar field.
The Lagrangian is invariant under
$$
\phi \rightarrow \phi ' = e^{i\alpha} \phi
$$
and hence the model has a $U(1)$ (global) symmetry. The vacuum
expectation value of $\phi$ is $\eta e^{i\alpha}$ where $\alpha$
can take any value. So the ground state of the model has 
continuous degeneracy. The degeneracy is labelled by the phase
angle $\alpha$ and hence the vacuum manifold is a circle.

Vortices are formed if we consider the model in two spatial dimensions
and let $\alpha$ be such as to wrap around the vacuum manifold. For
example, we could take $\alpha =\theta$, the polar angle.
Then, since the field is single valued everywhere, there
must be at least one point at which $\phi =0$. The field carries
energy at this point since $\phi =0$ is not on the vacuum manifold.
The location of this point may be defined as the location of a vortex.
(An example of a vortex is universally encountered by people taking
baths or washing dishes. As the water flows down the drain, it
circulates. We cannot interpolate the circulating velocity field 
all the way to the center of the vortex since it would
have to become multi-valued. Instead the fluid density in the central
region of the vortex vanishes.)

If we now take the model in three spatial dimensions, the vortex
becomes a line stretching in the third dimension. The vortex line
is called a ``string''.

The crucial element in the existence of the vortex was that $\alpha$
could ``wrap around'' the vacuum manifold. In other words, vortices
exist if the vacuum manifold contains incontractable closed paths.

Bogomolnyi's method cannot be applied to construct the vortex
solution in model (\ref{glstring}). In fact, the energy of an isolated
global vortex diverges. If the model (\ref{glstring}) is gauged, then
Bogomolnyi's method does work for a particular choice of parameters
in the model. Note that gauging the model makes no difference to the
vacuum manifold and so the topological arguments that show the 
existence of vortices still apply.

\subsection{Monopoles}

The model:
$$
L = |\partial _\mu {\vec \phi} |^2 - {\lambda\over 4}
                        ({\vec \phi}^2 -\eta^2 )^2
$$
where $\vec \phi$ is a triplet of scalar fields contains global monopole
solutions. To see this, note that the Lagrangian is invariant under
$$
{\vec \phi} \rightarrow {\vec \phi} ' = {\bf R} {\vec \phi}
$$
where, ${\bf R}$ is a rotation matrix in three dimensions.
Hence the model has an $O(3)$ (global) symmetry which is broken
down to $O(2)$ once $\vec \phi$ gets a vacuum expectation value.
For example, if $\vec \phi  \propto {\hat e}_3$, then the rotations
in the unbroken group are the rotations about the ${\hat e}_3$ axis. 

A monopole solution is obtained when
$$
\vec \phi \propto {\hat e}_r = ({\rm sin}\theta {\rm cos}\phi , 
{\rm sin}\theta {\rm sin}\phi , {\rm cos}\theta )
$$
where $\theta$ and $\phi$ are the angular spherical coordinates.
For this ``hedgehog'' configuration of the field, there must be 
a point in space where $\vec \phi =0$ and the energy density is
non-vanishing. In fact, in the global symmetry case the energy
of the monopole is infinite because of the slow fall off of gradient
energy at infinity. If the model is gauged, the $\vec \phi$ field
configuration can be accompanied by a gauge field that cancels off
the gradient energy at infinity. This then leads to a finite energy
solution but with a non-vanishing magnetic flux at infinity - the
famous ``magnetic monopole'' of `t Hooft and Polyakov \cite{th,ap}.

For the magnetic monopole, the field on the asymptotic two sphere
has to be non-trivial. So if the vacuum manifold admits incontractable
two spheres, we can have mappings from spatial infinity to the
vacuum manifold that cannot be smoothly deformed to the trivial
mapping (in which all of space is assigned the same point on the
vacuum manifold). And each such mapping would lead to a monopole
solution.

\subsection{Textures}

A simple model with (global) texture is 
\begin{equation}
L = (\partial _\mu \phi^a )^2 - 
                {\lambda\over 4} (\phi^a \phi^a -\eta^2 )^2
\end{equation}
where $a=1,2,3,4$. Now the vacuum manifold is a three sphere. If
our universe is a three sphere, $\phi^a$ could wrap around it an
integer number of times. Such a solution is a texture - the field
does not vanish anywhere but there is still some gradient energy
present.

In the cosmology literature, the term texture does not necessarily
refer to the case when the universe is a three sphere. One can 
consider a ball in space and find the winding of the field within 
this ball. Such configurations can have any winding. The term texture
is taken to apply to the situation where the winding is unity. Such
textures are time-dependent solutions. They collapse to a point where
the configuration unwinds and then the energy radiates away. (The 
scalar field is zero at one point in space-time.) 

\subsection{Hybrids}

In a sequence of symmetry breakings, one can get ``hybrid''
defects such as walls bounded by strings, or, monopoles connected
by strings. For example, in the symmetry breaking sequence:
$$
SU(2) \rightarrow U(1) \rightarrow 1
$$
the first symmetry breaking yields monopoles which get connected
by strings in the second stage of symmetry breaking.
Similarly one can get domain walls that are bounded by strings.
For further discussion of hybrid defects see Vilenkin and 
Shellard \cite{avps}.

\subsection{Topological Criterion and Homotopy Groups}

The criterion for having domain walls in a model in which
the symmetry group $G$ is spontaneously broken to $H$ by
the vacuum expectation value of a field can now be specified.
First, if $H$ is the trivial group, any element of $G$
acting on the order parameter would yield a possibly new
value of the order parameter which would still be in the minima
of the potential and hence would be a degenerate vacuum. Then the
manifold of vacuum states is simply given by the manifold
of $G$. Next, if $H$ is not the trivial group, the non-trivial
elements of $H$ leave the order parameter invariant. But since
$H$ is a subgroup of $G$, there are elements of $G$ whose action
on the order parameter is identical. In fact, the elements of $G$ 
can be ascribed to equivalence classes - the action of each element
within an equivalence class on the order parameter is identical but
the action of two elements belonging to two different equivalence 
classes can be different.
These equivalence classes are nothing but the elements $gH$ of the
coset space $G/H$ - elements of $gH$ acting on the order parameter
give the same result as $g$ acting on the order parameter since
elements of $H$ leave the order parameter invariant.
So the vacuum manifold is the manifold corresponding to $G/H$. 
The connectivity of a manifold is described by the zeroth
homotopy group of the manifold: $\pi_0 (G/H)$.
Domain walls are present in the model if the vacuum
manifold has disconnected components, that is, if 
$\pi_0 (G/H)$ is not trivial.

Now we can generalize these considerations further. The next
step is to consider a vacuum manifold that contains incontractable 
closed paths. For example, the vacuum manifold could be a one
sphere as in the $U(1)$ case when $\phi$ is a complex scalar
field. Then the vacuum expectation value of $\phi$ at infinity
determines a path in the vacuum manifold (and vice versa) by
the equation:
$$
\phi_\infty (\theta ) = g(\theta ) \phi_\infty (0)
$$
where $\theta \in [0,2\pi ]$ is the polar angle. The group
elements $g(\theta )$ determine a path on the vacuum manifold
parametrized by $\theta$.
Then one could consider a configuration $\phi_\infty (\theta )$ 
for which $g(\theta )$ traces one of the incontractable closed paths.
This mapping from the circle at spatial
infinity to the vacuum manifold has a non-trivial topological
index and hence the mapping from any two dimensional disk bounded 
by the circle at spatial infinity to the vacuum manifold must have a
singularity. In the field theory case, the singularity is simply
a location where the field $\phi$ vanishes. Now, since we can
choose to look at any surface bounded by the circle at spatial 
infinity, there must be a curve on which $\phi$ vanishes. 
The form of the potential tells us that there is energy wherever
$\phi =0$ and so there is energy distributed along a one-dimensional
curve - that is, a string. The conclusion is that there are strings
in the model whenever there exist incontractable paths on the vacuum
manifold. 

In mathematical language, paths on manifolds are put in equivalence
classes depending on whether they can be deformed into one another (that
is, an equivalence class contains homtopically equivalent paths).
Furthermore, there are combination rules for paths - two paths can be
combined to give a third path. The set of homotopically equivalent
paths, together with the combination rule for paths leads to a group
structure which is called the first homotopy group. For the manifold
$G/H$, this is denoted by $\pi_1 (G/H)$. In this language, the field 
theoretic model contains strings whenever $\pi_1 (G/H)$ is non-trivial.

The next generalization is to consider vacuum manifolds on which there 
exist incontractable two spheres. Here the mapping from the two sphere 
at infinity to the vacuum manifold can be non-trivial and this would
lead to a point-like singularity which we call a ``monopole''. 
Therefore there are monopoles in the model if $\pi_2 (G/H)$ is
non-trivial.

One could also consider the possibility of incontractable three spheres 
on vacuum manifolds. These would be given by non-trivial third 
homotopy groups, $\pi_3 (G/H)$, and result in ``textures''~\footnote{The
use of the word ``texture'' in condensed matter is different - it
refers to any variation of the order parameter, not necessarily having
non-trivial $\pi_3$.}. If space is a three sphere, we can have
a texture wrapped around the entire three sphere and this would give
a static solution in the model. 

The problem of finding the types of topological defects present
in a given model reduces to finding the homotopy groups for
a certain symmetry breaking $G \rightarrow H$. That is, we need to
find $\pi_n (G/H)$ ($n=0,1,2,3$) given the groups $G$ and $H$. In general, 
this can be quite complicated but there is an immensely useful theorem which 
is often applicable and simplifies matters. I now describe this theorem.

\begin{figure}[tbp]
\epsfxsize = 4 in \epsfbox{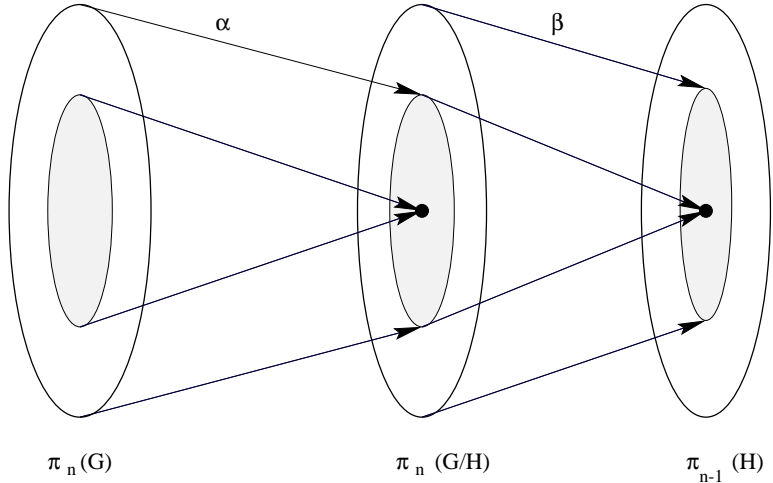}
\caption{An illustration of a part of the homotopy sequence.
The largest ellipse denotes the homotopy group specified below the 
ellipse, the middle shaded ellipse denotes the kernel of the map which is 
also the image of the previos map, and the innermost blackened ellipse 
denotes the identity element.\label{fig:sequence}}
\end{figure}

Mathematicians have found that there are homomorphisms between certain
homotopy groups that can be written as an ``exact homotopy sequence''. 
An important sequence is:
$$
... \rightarrow \pi_n (G) \stackrel{\alpha}{\rightarrow}
                \pi_n (G/H) \stackrel{\beta} {\rightarrow}
                \pi_{n-1} (H) \stackrel{\gamma} {\rightarrow} 
                \pi_{n-1} (G) \rightarrow ...
$$
The sequence denotes that there exist homomorphisms $\alpha$,
$\beta$ and $\gamma$ such that
$$
{\rm Im}(\alpha )= {\rm Ker} (\beta )
$$
$$
{\rm Im}(\beta )= {\rm Ker} (\gamma )
$$
where, Im() stands for the image of a map and Ker() stands for the
kernel of a map (i.e. the set of all elements mapped to the identity
element). This relationship is depicted in Fig. \ref{fig:sequence}.

In particular, if $\pi_n (G)$ and $\pi_{n-1} (G)$ are trivial 
(that is, contain the identity element only), ${\rm Im}(\alpha )$
has to be the identity of $\pi_n (G/H)$. 
But then, by the exact sequence, ${\rm Ker}(\beta )$ is also
the identity element. Using this fact one can prove \cite{herstein} 
that $\beta$ must be one to one. Next, we show that 
$\beta$ is onto. The homomorphism $\gamma$ in the sequence 
maps the entire group $\pi_{n-1} (H)$ to $\pi_{n-1} (G)$ since 
$\pi_{n-1} (G)$ is trivial. That is, ${\rm Ker}(\gamma ) =\pi_{n-1}(H)$.
Then ${\rm Im}(\beta ) = {\rm Ker}(\gamma ) =\pi_{n-1}(H)$ and
so $\beta$ is both one-one and onto and hence, is an isomorphism. 
Therefore:
$$
\pi_n (G/H) = \pi_{n-1} (H)
$$
if $\pi_n (G) = 1 = \pi_{n-1} (G)$.

As an example, for $n=2$, we know that $\pi_2 (G) =1$ for any
Lie group and, if $\pi_1 (G) =1$, then
$$
\pi_2 (G/H) = \pi_1 (H) \ .
$$
In particular, if $H$ contains any $U(1)$ factors, $\pi_1 (H)$ is non-trivial
and the model has monopoles. In Grand Unified Theories, the Grand Unified 
group is usually taken to be simply connected and then, since $H$ necessarily 
contains the electromagnetic $U(1)$ symmetry group, $\pi_2 (G/H)$ 
is non-trivial and hence, (magnetic) monopoles are necessarily predicted.

There is a subtlety in the homotopic classification that we
have glossed over. The subtlety is that the first homotopy
group classifies the paths on the vacuum manifold that pass
through some arbitrary but fixed base point. It is possible
that two closed paths may not be deformable to each other
if we impose the constraint that they should continue to 
pass through the base point but, if we were to relax this
constraint, they would indeed be deformable into one another. 
As there 
are no constraints on the field configurations that correspond 
to having a fixed base point, strings are classified by ``free'' 
homotopy. This subtlety does not play a role when the first
homotopy group is abelian but it can be important when the
group is non-abelian.

This subtlety also applies in a slightly varied form to the 
classification of magnetic monopoles. It can be important in
the cases where both strings and monopoles are present in
the model \cite{avps}.

\subsection{Exceptions: Semilocal Strings}

The classification of defects by homotopy groups is purely
topological and does not take any dynamical effects into
consideration. For example, the first homotopy group tells
us that there are non-trivial paths on the vacuum manifold
but does not tell us which of these paths will be preferred
when we solve the field theoretic equations of motion. For the 
same reason, there are a number of solitons that cannot be detected
by the homotopy classification. These include non-topological
solitons and semilocal strings \cite{tvaa,mhsemi,jpsemi}. I will discuss 
the latter here.

Consider the Lagrangian 
$$
L = |D_\mu \Phi |^2 - {1\over 4} F_{\mu \nu} F^{\mu \nu}
     - {\lambda \over 4}( \Phi^{\dag} \Phi - \eta^2 ) ^2
$$
where $\Phi$ is an $SU(2)$ doublet, and,
$$
D_\mu = \partial_\mu - ieA_\mu  
$$
where, $A_\mu$ is an Abelian gauge potential.

Note that the model has an $[SU(2)\times U(1)]/Z_2$ symmetry
where the $SU(2)$ is a global symmetry and the $U(1)$ is
a local (gauged) symmetry. (The $Z_2$ in the denominator
is present because the center of the $SU(2)$ factor is also 
contained in the $U(1)$ factor. Without modding out by $Z_2$,
we would be including these elements twice.) Once $\Phi$ acquires 
a vacuum expectation value, the symmetry is broken to $U(1)'$
which is a global symmetry group. This model is exactly 
the bosonic sector of the standard model of the electroweak 
interactions with weak mixing angle $\theta_w = \pi /2$.

The vacuum manifold of the standard model is known to be
a three sphere. The simplest way to see it is to consider
the minima of the potential:
$$
\Phi ^{\dag} \Phi = \eta^2
$$
and this is a three sphere. The homotopy classification 
now tells us that the three sphere is simply connected and
so there are no strings in this model. But this conclusion
is wrong.

A way to understand why there may be strings in the model
is that if one only considers the gauged symmetries, then
the breaking pattern is $U(1) \rightarrow 1$. And we know
that this symmetry breaking yields strings. However, since
the string is not topological, there are ways to deform the
string so that it is equivalent to the vacuum. What might
prevent this from happening is not topology but an energy 
barrier. A stability analysis for semilocal strings tells
us that they are stable only if $\lambda / 2e^2 < 1$.

\section{Formation in Cosmology}

Now we are ready to discuss the cosmological
formation of topological defects.

\subsection{General Discussion}

In a symmetry breaking, the order parameter
$\phi$ can have a vacuum expectation value
anywhere on the vacuum manifold. For example,
in the case of a $U(1)$ symmetry breaking, at
some time after the symmetry breaking,
$\phi = \eta {\rm exp}(i\alpha )$ where $\alpha$
can be any space dependent function. We expect
that the thermal nature of the symmetry breaking
would lead to a domain structure of $\alpha$
where the correlation of $\alpha$ at two 
spatial points decays with the distance between
the two points. In addition, large variations of
$\alpha$ should be suppressed by the terms in the
effective action that contain the gradient of
$\alpha$. In other words, we expect 
$$
| \nabla \alpha | \sim T
$$
that is, $\alpha$ changes by order 1 as a distance
of $\sim T^{-1}$ is traversed~\footnote{More rigorously
one should find the correlation length at the phase
transition and use that to find the $\alpha$ domain size.
This will be discussed in the lectures by Professor Kibble.}.

While the above argument is reasonable for the breaking
of global symmetries where $\alpha$ is a physical variable, it 
breaks down when the symmetries are gauged \cite{rudazam}. The point 
here is that the gradient of the phase of $\phi$
has no gauge invariant meaning. The gradient of $\alpha$
must now be replaced by the covariant gradient which 
also contains the gauge fields in it. Now we have
$$
| \nabla \alpha - e A | \sim T \ .
$$
This means that we can transfer some of the variation
in $\alpha$  to the gauge field. But the defect depends
on 
\begin{equation}
\oint d\alpha
\label{winding}
\end{equation}
where the integral is taken
along a closed path. This closed line integral is gauge invariant
but a relation with the covariant derivative - which is the 
physical one - cannot be made without first specifying the
gauge field. 

The bottom line is that one cannot say that $\alpha$ has
a certain value at a spatial point unless one fixes the
gauge. In the lattice simulations that I will describe, this 
difficulty is ignored and it is assumed that $\alpha$ can be treated 
as being a physical field just as in the global symmetry case.
In more recent work, this assumption has been questioned but
further work has provided some justification for working with 
$\alpha$ as if it was physical.

\section{Lattice Simulations}

\subsection{Domain Walls}

The simplest domain walls are formed when the
vacuum is two-fold degenerate. Let us call these
vacuua + and -. At the phase transition, there are
domains in which the order parameter chooses the
+ vacuum and other domains in which it chooses the
- domain. Since these are degenerate, the probability
for choosing + or - is 1/2.

What is the size distribution of walls formed in this
phase transition?

This problem is closely related to the problem studied in
percolation theory. There the problem is to assign
+ with probability $p$ and - with probability $1-p$
to every domain and then to find out the cluster size
distribution of + domains. The result is that, if
the domains are taken to form a cubic lattice, there
is one infinite size, + domain cluster if $p > 0.31$.
(That is, the + domains percolate if $p > 0.31$.)
In our problem, $p=0.5$ and so both + and - will
percolate. The boundary of + and - domains
is the location of domain walls and so
we will get one infinite domain wall. In addition, we
might find a few smaller domain walls.

The cluster size distribution is easy to find by doing
a simulation. In Table 1, I give the results from a
paper by Vilenkin and me \cite{tvav}.

\begin{table}[t]
\caption{
Size distribution of + clusters found by simulations on a
cubic lattice. 
}\vspace{0.4cm}
\begin{center}
\begin{tabular*}{12.0cm}{|c@{\extracolsep{\fill}}ccccccc|}
\hline 
{Cluster size} & {1} & {2} & {3} & {4} & {6} & {10} & {31082} \\[0.20cm]
\hline 
{Number} & {462} & {84} & {14} & {13} & {1} & {1} & {1} \\[0.20cm]
\hline
\end{tabular*}
\end{center}
\end{table} 

\subsection{Strings}

The formation of strings can be studied numerically
by assigning the phase $\alpha$ randomly on lattice
sites - say of a cubic lattice. Then one can go along
the edges of each plaquette of the lattice and evaluate
differences in $\alpha$. To do this, it is necessary to 
interpolate between the values of $\alpha$ at two sites.
Then one finds the integral in eq. (\ref{winding}) around 
a plaquette. If this is non-zero, it indicates
that there is a string or anti-string passing through
the plaquette. In this way, all the strings are found.
Then they are connected and information about the
distribution of string is stored.

\begin{figure}[tbp]
\epsfxsize = \hsize \epsfbox{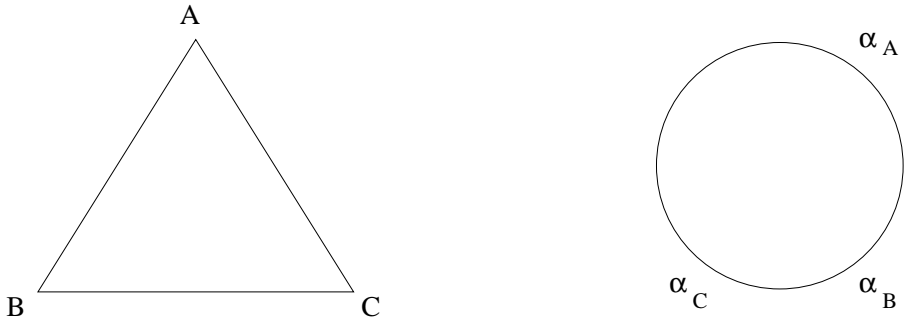}
\caption{The triangle is a plaquette in space and the circle denotes
the vacuum manifold. At each vertex of the plaquette a phase is 
assigned at random. In traversing from $A$ to $B$ on the triangle,
the phase must change from $\alpha_A$ to $\alpha_B$. However, there
is an infinite degeneracy in the path from $\alpha_A$ to $\alpha_B$
on the vacuum manifold since the path can wrap around the entire
circle any number of times.
\label{fig:algo}}
\end{figure}

A subtle issue in calculating the integral above is the interpolation 
as we go from one lattice site to another. Consider a $U(1)$ string 
simulation as shown in Fig. \ref{fig:algo}. 
As we traverse the triangle ABC in space,
the phase varies from $\alpha_A$ to $\alpha_B$ to $\alpha_C$ and then 
back to $\alpha_A$. These are simply points on a circle and we know that 
there are infinitely many paths joining any two points on a circle. 
(The paths can go around the circle infinitely many times.) So at every 
stage of the construction, we need to interpolate between the phases and 
there is an infinite-fold ambiguity in this interpolation. 
How do we resolve this ambiguity?

In the case of global strings, it is assumed that the shortest
of the infinitely many paths is the correct one. The rationale for
this choice is that the free energy density gets contributions from
a term $|\nabla \alpha |^2$ and this is least for the shortest
path. The rule of choosing the shortest path to interpolate 
between two points on the vacuum manifold is known as the 
``geodesic rule''.

In the case of gauge strings, the rationale for the geodesic
rule breaks down since the contribution to the free energy
involves the covariant derivative of $\alpha$ and not the
ordinary derivative. Now which path should be chosen?
Following the logic of the global case, it should be the
path that minimizes $|\nabla \alpha - e A|^2$. In the simulation
this would mean that we should not only keep track of $\alpha$
but also the gauge field $A$. We will discuss a possible way
to circumvent this problem in the next subsection while here
we will assume the geodesic rule to be valid. 

The surprising result that emerges from numerical simulations
is that most of the energy in the string network is 
in infinite strings. Furthermore, the strings are Brownian on
large scales and the loop distribution is scale invariant. Let
us explain these results in more detail:

\begin{itemize}
\item Brownian strings: This means that the length $l$ of a string is
related to the end-to-end distance $d$ by
\begin{equation}
l = {{d^2} \over \xi}
\label{brownian}
\end{equation}
where $\xi$ is a length scale also called the step length which would
be roughly given by the lattice spacing. This result is
valid for large $l$. For smaller lengths, the walk is not Brownian
and lattice effects are also present. 

\item Loop distribution: Scale invariance means that there is no
preferred length scale in the problem apart from the lattice cut-off
and loops of all sizes are present. So the number density of loops
having size between $R$ and $R+dR$ is given by dimensional analysis:
$$
dn(R) = c {{dR} \over {R^4}}
$$
where $c\sim 6$. Using eq. (\ref{brownian}), this
may be written as:
$$
dn(l) = {{c} \over {2\xi^{3/2}}} {{dl}\over {l^{5/2}}} \ .
$$
Note that the scale invariance is in the {\it size} of the loops
and not in their {\it length}. 

\item Infinite strings: With the implementation of the geodesic
rule, the density in infinite strings was estimated
to be about 80\% of the total density in strings. The way this
estimate was made was to do the simulation in bigger and bigger
lattices and keep track of the length in the strings that were
longer than a large (compared to the lattice size) critical 
length \cite{tvav}. 
As the lattice was made bigger, the fraction of string in long strings 
tended to stabilize around 80\%. Simulations on other lattices and with
periodic boundary conditions also yield infinite strings but the
estimated fraction can vary upward from about 74\%. Analytic estimates 
of the fraction which assume that the strings are random walks on a lattice, 
are consistent with these estimates \cite{scherrerfrieman}.

\end{itemize}

Can one analytically see the presence of infinite strings?
This is an open question. Some progress can be made if one assumes that 
strings perform a Brownian walk \cite{scherrerfrieman}. It is known that 
random walks do not close in 3 dimensions and this tells us that infinite 
strings will be present. Furthermore, estimates can be obtained for
the fraction of length in infinite strings and the result is similar 
(though not identical) to the one obtained by simulations.

For cosmological applications such as the formation of large-scale
structure, the existence of infinite strings is vital. The reason
is that the small closed loops can decay by emitting gravitational 
and other forms of radiation but the infinite strings are destined
to live forever because of their topological character~\footnote{The 
two ways in which they could decay are: a) a string meets an antistring
and annihilates, and, b) a string snaps leading to a gravitational
singularity. Neither process is expected to occur at a rate that
would be cosmologically interesting.}. So only the infinite strings (and
their off-spring loops) could live to influence late time cosmology and
also to tell us the story of the Grand Unified epoch~\footnote{Professor 
Kibble has raised the interesting possibility that a population of long 
loops might be able to play the role of infinite strings in cosmology.}.

\subsection{Relaxing the Geodesic Rule}

A possible cure for the ambiguity in choosing the path on the 
vacuum manifold (discussed above) is to relax 
the geodesic rule and
assume that the phase difference between two lattice sites is given
by a probability distribution \cite{lptv}.
So, if the values of the phases at lattice sites 1 and 2 are
$\theta_1$ and $\theta_2$, the phase difference will be
$$
\Delta \theta = \theta_2 - \theta_1 + 2\pi n 
\equiv \delta \theta + 2\pi n
$$
where $n$ is a random integer drawn from some distribution. A
convenient choice for the distribution is
\begin{equation}
P_n = \int_{n-0.5}^{n+0.5} dm
2 \sqrt{\pi \beta} e^{-\beta (\delta \theta + 2\pi m)^2} \ .
\label{pdist}
\end{equation}
with $\beta > 0$ being a parameter.
This probability distribution is consistent with the idea
that longer paths on the circle should be suppressed but the
amount of suppression depends on the choice of $\beta$. Note
that $\beta$ plays the role of inverse temperature since
lower values of $\beta$ (that is, higher temperatures) allow
for larger values of $n$ while larger values of $\beta$ reduce
the algorithm to the geodesic rule.

\begin{figure}[tbp]
\centerline{
\epsfxsize = 2.25 in \epsfbox{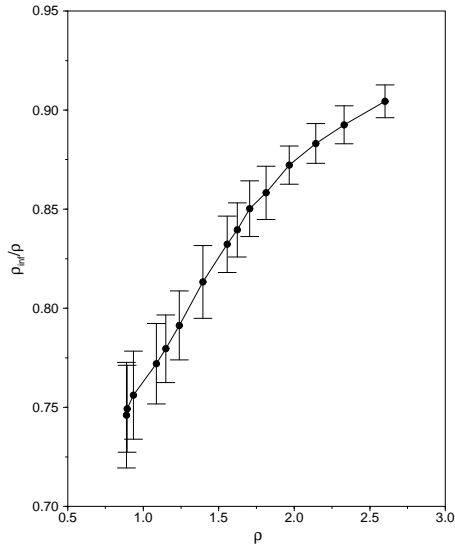}
}
\vskip 1.5 cm
\caption{A plot of the infinite string density fraction versus the total 
string density. The total string density increases as the parameter 
$\beta$ is lowered. The geodesic rule is recovered
in the limit that $\beta$ becomes very large.
\label{fig:relax}}
\end{figure}

In simulations that relax the geodesic rule \cite{lptv}, it is found
that the fraction of infinite strings gets larger with smaller values 
of the parameter $\beta$. And hence the case for having infinite strings
in a realistic phase transition is strengthened (see Fig. \ref{fig:relax}). 

\subsection{Monopoles}

Lattice simulations of monopoles largely follow the strategy adopted for 
strings \cite{leeseprokopec}. Here too one throws down random phases 
(corresponding to points on a two sphere) on lattice points, adopts the 
geodesic rule, and then finds the winding of the configuration. The 
calculations are a little more involved since one needs to calculate windings 
of a two sphere on a two sphere. Also, a cubic lattice leads to ambiguities 
and it is better to work on a lattice where the cells are (irregular) 
tetrahedra. 

The results of these simulations is a distribution of monopoles
which is correlated on small scales - for example, the cells 
sharing a plaquette with a cell containing a monopole can only 
contain an antimonopole - but these correlations decay with distance. 
The distance between closest
monopoles is not too much larger than the distance between closest
monopole and antimonopole \cite{tvbias}. As a result, if the system is 
left to evolve further under the mutual forces of the monopoles, some of
the close-by monopoles and antimonopoles annihilate but very often
a monopole is left as an ``odd man out''. The antimonopole it could
annihilate with is located far away and was also the ``odd man out''
in a set of local annihilations. As a result, the correlations between
monopoles and antimonopoles are quickly washed out and the end
result is a scaling density of poles~\footnote{These results on the 
evolution of monopoles were derived by Preskill \cite{preskill} and 
have been confirmed numerically \cite{leeseprokopec}.}.

\subsection{Biased Topological Defects}

So far we have assumed that the vacuum manifold is completely
degenerate. But in many circumstances, the degeneracy may be
broken by explicit symmetry violations. For example, in the
domain wall case, the + vacuum might be slightly preferred
over the - vacuum. This is likely to be relevant in condensed
matter physics where there are external forces present ({\it eg.}
earth's magnetic field, rotation of the container etc.) that
could ``bias'' the symmetry. What happens to the statistics of the
defects in such cases?

In the case of domain walls, as already discussed, the result is 
known from percolation theory. At a critical bias, the probability
of a domain having + falls below 0.31 (on a cubic lattice) and the
+ domains do not percolate. Then all the domain walls are finite.

In the case of strings, we do not have results from percolation
theory but a few simulations have been done \cite{tvbias,mhstrobl}. 
It is found that the infinite strings break down into smaller loops at 
some critical value of the bias and the resulting distribution of loops 
is not scale invariant. Instead it is well fitted by:
$$
dn(l) = {{c} \over {2\xi^{2}}} {{dl}\over {l^{a}}} 
              {\rm exp}[-b l] \ 
$$
where $a, b$ and $c$ are constants that depend on the value of the
bias. In particular, note that the exponent $a$ need not be the scale 
invariant value of $5/2$. Instead two regimes seem to be indicated -
for large bias $a=2$ while for small bias $a=5/2$.

The feature that I found amusing in the simulation of biased defects was 
that the transition from percolating strings to non-percolating strings 
appears to be rather sudden. In other words, there seems to be a critical 
minimum amount of bias that prevents infinite strings from 
forming~\footnote{I sometimes wonder if this work can have
application in the real world. Suppose that a manufacturer produces
metal sheets by cooling molten material but is worried about the line
defects that would form during cooling, run across the sheets and weaken 
them. One way to get rid of the line defects would be to use some
sort of bias while cooling. But the question is if there is an optimal 
amount of bias that will be enough to clean the sheets from (long) line 
defects or whether it is simply that the more the bias, the fewer defects 
are produced. What we find here is that an optimal bias indeed exists at 
which the infinite line defects that run across the sample disappear even 
though the exact value of the bias will depend on the type of line defect, 
type of bias and other details.}.

Bias can also be added to simulations of monopoles. The relevant
quantities to study here are the distance between closest monopoles
and the distance between closest monopole ($m$) and antimonopole ($\bar m$), 
as functions of the bias. In the simulations it is found that the
$mm$-distance grows with bias but the $m\bar m$-distance stays
roughly constant. At some stage the monopole distribution is in 
monopole-antimonopole pairs with different pairs being widely separated.
In this case, further evolution would cause the paired monopoles and the
antimonopoles to annihilate and their would be no confusion about
which monopole should annihilate which monopole. All the monopoles
would then disappear.

\subsection{Problems with Lattice Based Simulations}

The simplest way to see that lattice based simulations might be 
suspect is to realize that the critical percolation probability 
depends on the lattice that is used. 

Consider the case of domain walls in which the
probability of laying down a + is $p$ and when the critical percolation
probability, $p_c$, is less than 0.5. Then there are three phases:
\begin{itemize}
\item  $p < p_c$: the + domains are islands in a sea of -, 
\item  $p_c <p< 1-p_c$: the + and - both form seas, and, 
\item  $1-p_c < p$: the + form the sea, the - form the islands.
\end{itemize}
In the unbiased case, we have $p=0.5$ and so we would always get
seas of + and -, and the boundary between the + and - regions would
also be infinite. That is, the domain walls are infinite in size.

If $p_c > 0.5$, the picture is quite different. Now we have:
\begin{itemize}
\item  $p < 1-p_c$: the + domains are islands in a sea of -,
\item  $1-p_c <p< p_c$: the + and - both form islands, and, 
\item  $p_c < p$: the - form islands in a sea of +.
\end{itemize}
Again, in the unbiased case, $p=0.5$ and so both the + and the
- form islands. The interface between the islands are finite in
extent and so there are no infinite domain walls.

\begin{figure}[tbp]
\vskip 2.5 truecm
\epsfxsize = 1.25 in 
\centerline{
\epsfbox{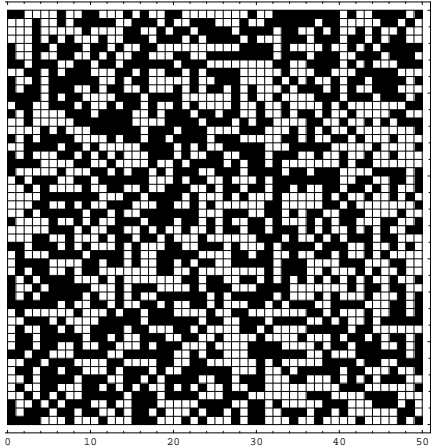}
}
\caption{The black squares denote + domains and the white squares
denote - domains in a simulation with $p=0.5$ on a two-dimensional
square lattice. Neither the + nor the - domains percolate in this 
case and the domain walls, which are the boundaries between black 
and white squares are all finite.\label{fig:sqlat}}
\end{figure}

So one sees that the value of $p_c$ as compared to $p$ is very
important for finding out if infinite domain walls are formed.
What is even more interesting is that, in two spatial dimensions,
$p_c=0.5$ for a triangular lattice and $p_c=0.59$ for a square 
lattice. So the domain walls in two dimensions with $p=0.5$ are 
(marginally) infinite on a triangular lattice and are all finite 
on a square lattice (see Fig. \ref{fig:sqlat}). 
Which lattice is the correct one to use to study phase transitions?

One expects the same problems to arise in the lattice based study of 
strings and monopoles. In fact, the study of domain walls is fundamental 
to understanding strings and monopoles since strings, for example, may be 
viewed as the intersection of two types of domain walls - one on which the 
real part of a complex scalar field vanishes and the other on which the 
imaginary part vanishes \cite{scherrervilenkin}. If the two types
of domain walls are all finite, the strings will also be finite.
Hence it is suitable to first understand the percolation of domain
walls.

\section{Lattice-Free Simulations}

It is easier to think about first order phase transitions since
here we know that bubbles of the new phase nucleate, collide, 
coalesce and eventually fill space with the new phase. At every
bubble collision there is a possibility that a defect will be
produced. So the first step is to study the processes that can
occur when two or more bubbles collide. This has been treated
analytically by Kibble and Vilenkin \cite{twbkav} and numerically 
by several 
groups \cite{bubcoll} mainly in the context of string formation.
It is found that bubble collision indeed leads to the formation
of strings. This can happen when two or more bubbles collide.
In the case of gauge strings, when two bubbles collide, a magnetic 
flux tube is formed in the shape of a closed ring at the location 
of the junction of the two bubbles. When several bubbles collide, 
these magnetic flux tubes can coalesce to form strings.

Numerical simulations have been performed to study the distribution
of vortices in two spatial dimensions \cite{jbtktvav} and strings in 
three spatial dimensions \cite{jb}. The three dimensional simulation 
indicates that there
may be an absence of infinite strings though the numerical limitations
do not permit any firm conclusion. For this reason, I will now
describe the easier task of studying domain wall formation in 
a first order phase transition in two spatial dimensions. The
extension to three spatial dimensions is conceptually straightforward
but has not yet been done.

\section{Percolation in First Order Phase Transitions}

\begin{figure}[tbp]
\centerline{
\epsfxsize = 2.65 in \epsfbox{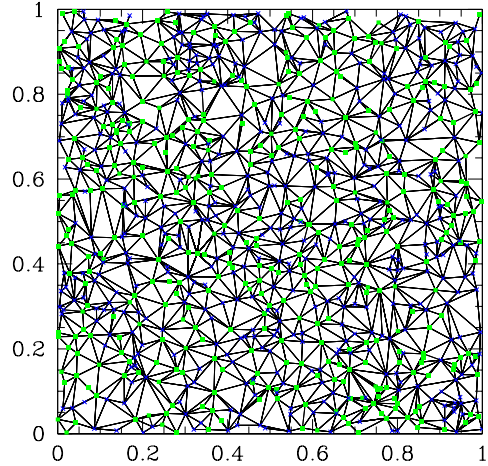}
}
\caption{The crosses denote bubble centers that are in the + phase
and the filled squares denote bubble centers that are in the - phase.
If two bubbles collide, their centers are joined by straight lines.
The figure then shows the ``bubble lattice'' expected in a first order 
phase transition in two spatial dimensions.\label{fig:bublat}}
\end{figure}

Imagine that a phase transition occurs in two spatial dimensions and
proceeds by bubble nucleation. Assume:
\begin{itemize} 
\item The bubble nucleation probability is constant per unit 
time per unit volume of false vacuum.
\item Once a bubble nucleates, its radius grows with constant velocity.
\item When bubbles collide they continue to grow as if there was no collision.
\end{itemize}
Already there are several interesting questions one can ask. For example,
what is the rate of bubble nucleation? How many collisions does an average
bubble have? These questions are important if one wants to connect percolation
in phase transitions to the previously studied percolation on a lattice.
Here I will not go into details but only mention that the average number of 
collisions of a bubble is about 6 and hence is very close to the number of
neighbors of a lattice site on a triangular lattice 
(see Fig. \ref{fig:bublat}). 

\begin{figure}[tbp]
\centerline{
\epsfxsize = 2. in \epsfbox{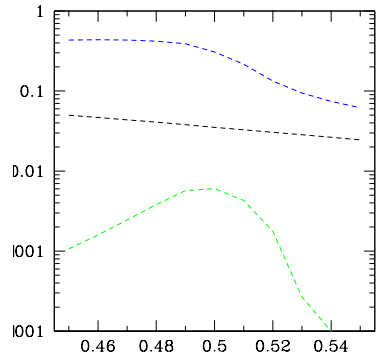}
}
\caption{The graphs show the moments of the + cluster size distribution
excluding the largest cluster for a range of probabilities, $p$. The top 
curve is the zeroth moment $\sum_s n(s)$ where $n(s)$ is the number density
of clusters having size $s$. The middle curve is $\sum_s s n(s)$ and the
bottom curve is $\sum_s s^2 n(s)$. From the bottom curve, it is clear that
the second moment turns over around $p=0.5$. This turnover marks the 
appearence of a very large cluster which is not included in the calculation
of the moments. Hence the critical percolation 
probability is 0.5. \label{fig:perc}}
\end{figure}

Percolation in the phase transition is studied by assigning a + or a -
to each bubble. If a + bubble collides with a - bubble, the two bubbles
are separated by a domain wall. We want to know the typical size of a
+ cluster for different probabilities, $p$, for assigning a + to a bubble.
The simulation results show that this probability is about 0.50 which is
the same as the result for a triangular lattice 
(Fig. \ref{fig:perc}). Hence we conclude 
that domain walls will (marginally) percolate in first order phase transitions 
in two spatial dimensions.

These results suggest what might happen in three dimensions and for
other defects. If the critical percolation probability for bubbles is
the same as that for tetrahedral lattices and walls percolate on a
tetrahedral lattice, we expect domain walls to
percolate in first order phase transitions in three
dimensions as well. Then it is very likely that the intersections of two
kinds of domain walls will also percolate - that is, infinite strings
will be present. It would be reassuring to confirm these suggestions in
actual numerical simulations.

While the above discussion lends confidence to the usual picture of
defect formation, it does not provide quantitative results into the
cluster size distribution and other features of interest. For this
an analytic treatment would be invaluable. However, at the moment
the subject is completely open and, to my knowledge, it is not even
known how to start to attack the problem.

\section{Directions}

As discussed above, lattice-based numerical simulations have been 
extensively used to study defect formation. These give a picture that
is well-defined and seems to hold up in general characteristics even
when modifications are made to make the simulations more realistic.

Lattice-free numerical simulations have just started and as far as we
can tell, have not changed the picture emerging from lattice-based
simulations in any dramatic way.

Field theoretic numerical studies have also just started and we have yet 
to see the picture that will emerge. These studies are very computer 
intensive but, as computer resources improve, could become valuable
for studying the formation of defects \cite{nalbmh}. They would also
contribute to our understanding of the formation of non-topological
defects such as semilocal \cite{aakklptv,aaetal} and electroweak 
strings \cite{yokoyama}.

The most notable shortage is in an analytical understanding of defect
formation. Here there is a lack of implementable techniques and there
is plenty of scope for improvement. It is also likely that certain issues 
have already been addressed in condensed matter physics and a search of
the literature may be a good starting point.

\section*{Acknowledgements}

This work was supported by the Department of Energy, USA.

\end{document}